\documentclass[11pt]{article}
\usepackage{moriond,epsfig}
\usepackage[latin1]{inputenc}
\usepackage[OT1]{fontenc}
\usepackage{amssymb}
\def\PRD{{\em Phys. Rev.} D}

\def\be{\begin{equation}}
\def\ee{\end{equation}}
\def\bea{\begin{eqnarray}}
\def\eea{\end{eqnarray}}

\begin{document}
\vspace*{4cm}
\title{BEYOND THE CHAMELEON MECHANISM}
\author{ D. F. MOTA$^1$ and D. J. SHAW$^2$}
\address{$^1$ Institut f\"ur Theoretische Physik, Universit\"at
Heidelberg,  D-69120 Heidelberg, Germany
\\ $^2$ School
of Mathematical Sciences, Queen Mary, University of London, London
E1 4NS, UK}
\maketitle
\abstracts{
As a result of non-linear self-interactions, in chameleon theories where the field
couples to matter much \emph{more strongly} than gravity does, the fifth force
between two bodies with thin-shell is \emph{independent} of their
coupling to the field. As a consequence the bounds on the
coupling coming from terrestrial
tests of gravity, measurements of the Casimir force and those
constraints imposed by the physics of compact objects, big-bang
nucleosynthesis and measurements of the cosmic microwave background anisotropies can be exponentially relaxed.}

Stringent experimental limits on the properties of light scalar fields
coupled to matter can be  relaxed if the scalar field theory in question
possesses a \emph{chameleon mechanism} \cite{c9}.  This
mechanism provides a way to suppress the forces mediated by
the scalar fields via non-linear field self-interactions.  A direct
result of these self-interactions is that the mass of the field is no
longer fixed but depends on the ambient density
of matter. The properties of these scalar fields therefore change
depending on the environment; for this reason such fields
have been dubbed \emph{chameleon fields}.
We consider  theories where the chameleon field, $\phi$, has a
self-interaction potential given by: $ V(\phi) = M^4
\left(M/\phi\right)^{n}$. 
If $M \sim 2 \times 10^{-3}\,{\rm eV} \approx (0.1\, \mathrm{mm})^{-1}$ the chameleon may play the role of dark energy \cite{chamcos}.
The equation of motion for $\phi$ is\cite{c9}
\begin{equation}
- \Box \phi = V_{,\phi}(\phi) + \beta\rho/M_{\rm pl}, \label{field}
\end{equation}
where $\rho$ is the energy density of matter and $\beta$ parametrizes the strength of the coupling to matter relative to that of gravity; $c = \hbar = 1; M_{\rm pl} = 1/\sqrt{8\pi G}$. The right hand side of Eq. (\ref{field}) vanishes
when $\phi = \phi_c(\rho)$: $$ \phi_c (\rho) =
M\left(\beta\rho/( n M_{\rm pl}M^3)\right)^{-\frac{1}{n+1}}. $$ For $\phi_c(\rho)$ to be
real when $\beta\rho > 0$, we need either $n \geq 0$
or for $n$ to be negative and even; and  $n\neq 0, -2$ for the theory to be non-linear. 
The mass of small perturbations about $\phi = \phi_c$ is $m_{\rm c} = \sqrt{V_{,\phi \phi}(\phi_c)} = \sqrt{
n(n+1)}M\left\vert M/\phi_c\right\vert^{n/2 +1}$. 

A body is
said to have a thin-shell if $\phi$
 is approximately
constant everywhere inside the body, except in a thin-shell near
the surface of the body where large changes in its value
occur.   Deep inside a body with a thin-shell
$\phi$ is constant, and so we might expect $\phi = \phi_{\rm c}(\rho)$.
The effective chameleon mass, $m_{\rm eff}$, in the body would then be
given by $m_{\rm eff}=m_c(\rho)$.  The thin-shell effect is in fact the reason why chameleons, which couple to matter with the similar strength as gravity does, $\beta \sim \mathcal{O}(1)$, 
evade experimental bounds on fifth forces and WEP violations. This effect 
suppresses the $\phi$-force between large objects. This is because experiments, in reality, probe only an effective coupling  given by:
$\beta_{\rm eff} = 3\beta\cdot \frac{\Delta R}{R}$. Where 
$R$ is the radius of an object and $\Delta R$ is the size of the (thin) region at the surface of the body 
where all variations in $\phi$ occur \cite{c9}.  However, the original analysis \cite{c9} was only valid for a coupling of gravitational strength,  $\beta \leq \mathcal{O}(1)$, and required that Eq. (\ref{field}) could be linearized. In fact, the non-linear nature of the potential plays actually a fundamental role. This is particularly apparent if one considers the strong coupling regime, $\beta \gg 1$, and   
%
becomes clear when determining the effective, large-scale behaviour
of the chameleon \cite{doug}. Eq. (\ref{field}) defines the
microscopic, or particle-level, field theory for $\phi$, whereas
in most cases we are interested in the large scale or coarse
grained behaviour of $\phi$. 
%
The effect of the non-linearities on
the averaging is to limit the averaged value of
$m_{\phi}$ to be smaller than some critical value, $m_{\rm crit}$
\cite{doug}. $m_{\rm crit}$ is a macroscopic quantity but it
depends only on the microscopic properties of the body and the
index $n$.  It is \emph{independent} of $\beta$ and $M$ \cite{doug}.  Modelling the body as being composed of
particles of radius $R_{\rm p}$ separated by an average distance
$d_{\rm p}$,  the macroscopic mass of the chameleon in the body can be shown to be \cite{doug}
$m_{\rm eff} = \min\left(m_{c}(\rho), m_{crit}\right)$, where: $$
m_{\rm crit} \approx \sqrt{3\vert n+
1\vert} d_{p}^{-1}\left(R_{p}/d_{p}\right)^{\frac{q(n)}{2}},
\, n \neq -4, $$ where $q(n) = \min(1, (n+4)/(n+1))$
.  Whenever $m_{\rm eff}=m_{\rm crit}$ is it because the individual particles that make-up the body have themselves developed thin-shells. This critical behaviour emerges from the requirement that non-linear effects are negligible outside of the particle from $r \gtrsim R_{p}$ to $r=d_{p}$: this implies a maximal value of $m_{\rm eff}$, i.e. $m_{\rm crit}$, that depends only on $R_{\rm p}$, $d_{\rm p}$ and $n$. The $n$ appears as it determines precisely when linearized theory breaks down\cite{doug}.

The $\beta$-independent critical behaviour is also seen in the $\phi$-force
between two bodies. The onset of this critical behaviour is linked to the
emergence of a thin-shell. A body of radius $R$ and density $\rho_{\rm c}$ in a
background of density $\rho_{\rm b} \ll \rho_{\rm c}$ has a thin-shell if:
\be m_{\rm eff}R \geq \sqrt{3\vert n+1 \vert}\left\vert 1-
\left(\rho_{\rm c}/\rho_{\rm b}\right)^{\frac{1}{n+1}}\right\vert^{1/2},
n \neq -4. \label{thinshell} \ee The existence of a thin-shell is
essentially due to non-linearities being strong near the surface
of a body but weak in other regions.  
When $n > 0$, $(\rho_{\rm c}/\rho_{\rm b})^{1/n+1} \gg 1$ and so the
thin-shell condition, eq.(\ref{thinshell}), depends greatly upon
on the background density. The same is \emph{not} true when $n
\leq -4$ since here $(\rho_c/\rho_b)^{(1/n+1)} \ll 1$.
Therefore $n > 0$ theories can behave differently in space-based
experiments than they do in laboratory ones, because the
thin-shell condition is more restrictive in low-density background
of space than it is in the lab\cite{doug}. 
In contrast, there will be no great difference between the predictions of $n \leq -4$ theories  for space and ground based tests.

The existence of a thin-shell in the test-masses used in experimental searches for deviations from general relativity is vital if we are to evade their bounds.
Whereas the force between two non-thin-shelled bodies with separation $r$ is $2\beta^2 (1+m_{\rm b} r)e^{-m_{\rm b} r}$ times the
gravitational force between them ($m_{\rm b}$ is the chameleon mass in the region between the bodies), the force between two
bodies, of masses $M_1$ and $M_2$, with thin-shells is found to be
independent of the coupling $\beta$ \cite{doug}. It is for this reason that strong couplings, $\beta \gg 1$, are allowed. When $d \gg R_1, R_2$, where
$R_1$ and $R_2$ are the respective radii of the two bodies, this
force is found to be $\alpha_{12}$ times the strength of gravity,
where for $n \neq -4$: $$ \alpha_{12} = \frac{S(n,m_b)
M_{\rm pl}^2(1+m_{\rm b} r)e^{-m_{\rm b} r}}{M_1 M_2} (M^2R_1R_2)^{q(n)}, $$ where
$S(n,m_{\rm b})$ is $(3/\vert n \vert)^{2/\vert n + 2 \vert}$ for $n <
-4$, whereas for $n > 0$ it equals $(n(n+1)M^2/m_{\rm b}^2)^{2/(n+2)}$.
This $\beta$-independence was previously noted for $\phi^4$ theory \cite{nelson}.

The $\beta$-independence can be understood as follows: just outside a
thin-shelled body, the $V_{,\phi}$ in eq. (\ref{field}) is large
and negative ($\sim O(-\beta \rho / M_{\rm pl})$), and so $\phi-\phi_{\rm b}$
decays very quickly. At some point $\phi-\phi_{\rm b}$ reaches a critical value,
$\delta\phi_{\rm crit}$, that is small enough so that non-linearities are no
longer important.  Since this all occurs outside the body,
$\delta \phi_{\rm crit}$ can only depend on the size of the body, the choice of
potential $(M,  n)$ and the value (and hence mass) of $\phi$ in the background.

This $\beta$-independence is of great of importance if one wishes to
design an experiment to detect the chameleon through WEP
violations. Since the $\phi$-force is independent of the coupling,
$\beta$, for bodies with thin-shells, any microscopic composition
dependence in $\beta$ will be hidden on macroscopic length scales.
The only `composition' dependence in $\alpha_{12}$ is through the
masses of the bodies and their dimensions ($R_1$ and $R_2$). 
If we measure the differential accelerations of two test
masses, $M_{1}$ and $M_{2}$, of radii $R_{1}$ and $R_{2}$ towards a
third body, mass $M_{3}$ and radius $R_{3}$, then the  E\"{o}tvos parameter is $\eta =
\alpha_{13}-\alpha_{23}$.  Taking the third body to be the Sun or the
Moon, experimental searches for WEP violations currently limit $\eta \leq 10^{-13}$. 
In most of these searches,
although the composition of the test-masses is different, they 
have the \emph{same} mass ($M_1=M_2$) and the \emph{same} size
($R_1 = R_2$).  Therefore, if the test-masses have thin-shells we have
$\eta = 0$ and \emph{no} WEP violation will be detected.  The only
implicit dependence of this result on $\beta$ is that the
\emph{larger} the coupling is, the more likely it is that the
test-masses will satisfy the thin-shells conditions. 
Hence, to detect a chameleon field through WEP violations, the test-masses must either not satisfy the thin-shell conditions or have different masses and/or dimensions.

\begin{figure}
\psfig{figure=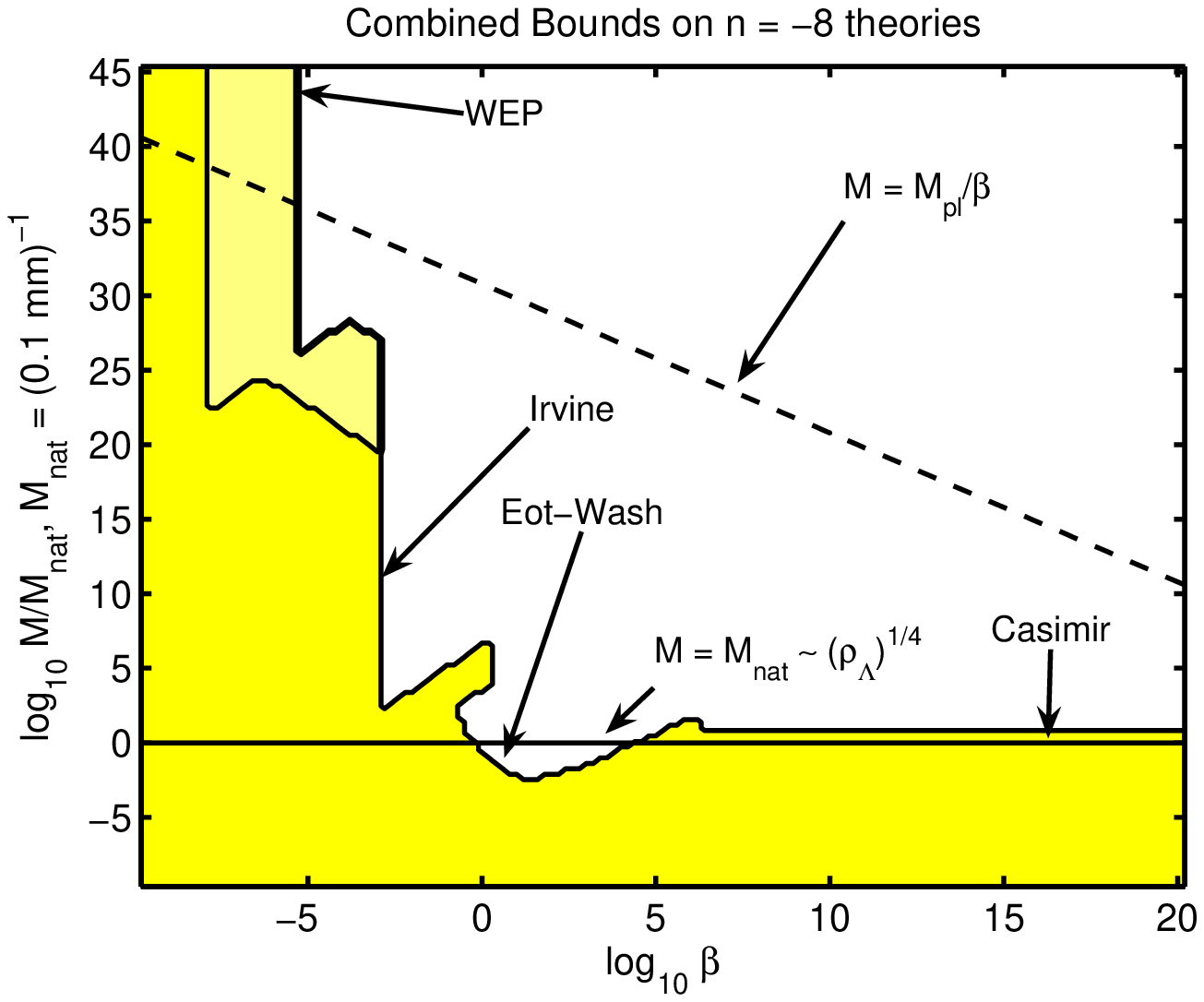,height=2.2in}
\psfig{figure=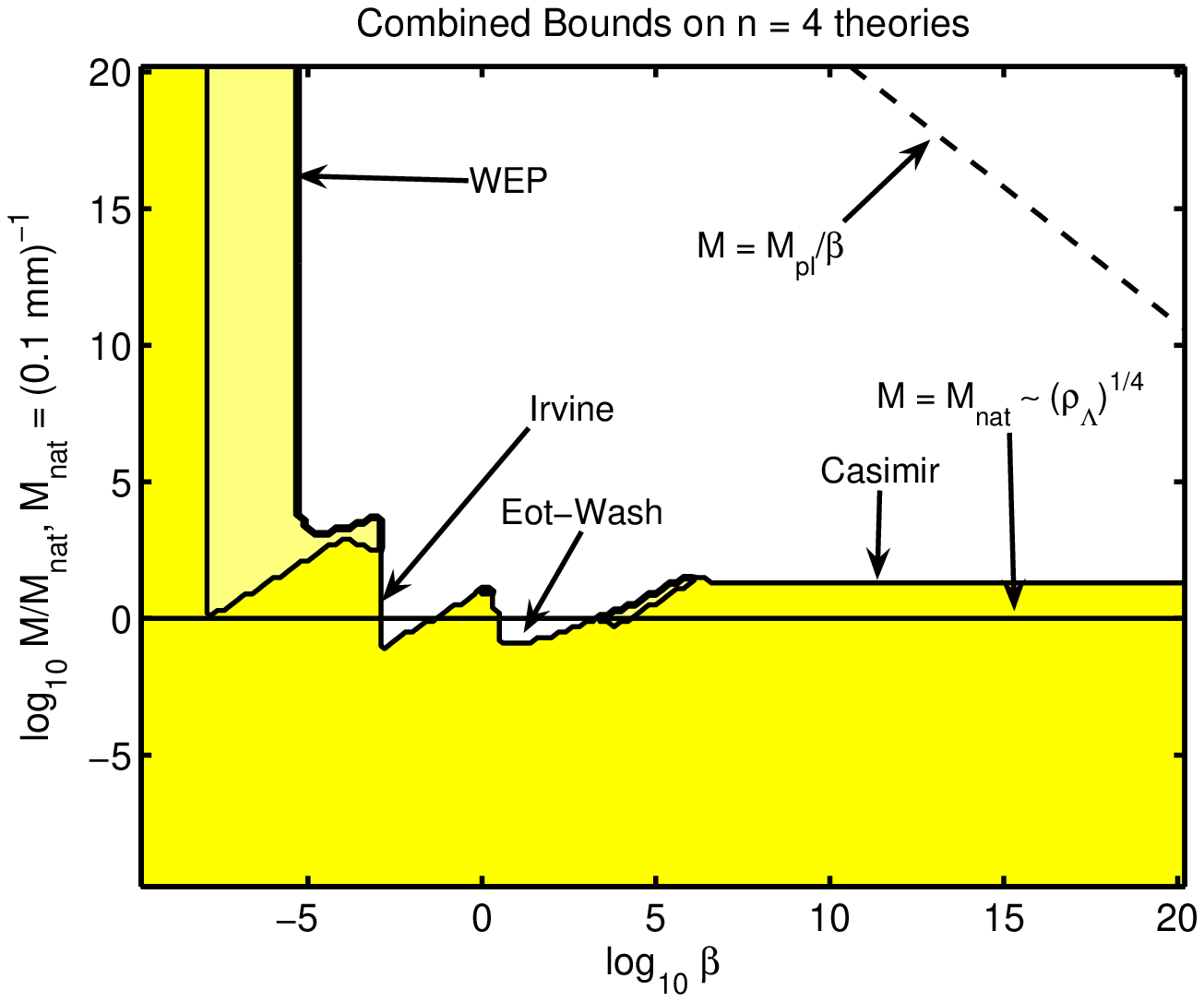,height=2.2in}
\caption{The whole of shaded area shows
the regions of parameter space that are allowed by the current
data.  Future space-based tests could detect the more lightly
shaded region.  The solid line indicate the cases
where the chameleon could be dark energy.  Dotted line indicates when 
chameleons 
would be produced in particle colliders.
\label{fig1}}
\end{figure}

Using two
spherical test bodies both with a mass of $10\,g$, where one is made
entirely of copper and the other of aluminium.  The strongest bounds on
chameleon fields would then come from measuring the differential
acceleration of these bodies towards the Moon.  We indicate in
FIG.\ref{fig1} the restrictions that finding $\eta \leq 10^{-13}$
in such an experiment would place on these chameleon theories.  
The
Moon is a better choice of attractor than the Earth or the Sun for
such experiments since $\alpha_{13}$ is proportional to
$M_{pl}^2/M_{1}M_{3}$ and so the smaller mass of the test-bodies,
$M_{1}$, and the attractor, $M_{3}$, the larger $\eta$ will be
compared to gravity.  The corollary of this result is that if we are
unable to detect $\phi$ in lab-based, micro-gravity experiments where
both $M_1$ and $M_2 \sim O(10 \, g)$ then the $\phi$-force between larger
objects, would also be undetectably small.  For this reason
measurements of the differential acceleration of the Earth and Moon
towards the Sun, e.g. lunar laser ranging, are not competitive with
lab-based experiments. Future, space-based tests of WEP promise to be able to detect $\eta$
up to a precision of $10^{-18}$; we indicate on FIG.\ref{fig1}, how such tests would improve the constraints.

The $\phi$-mediated force will also produce effective corrections to
the $1/r^2$ behaviour of gravity, which are constrained by the E\"{o}t-Wash experiment which probes gravity over  separations $d \geq 0.1 \mathrm{mm}$, with the bound being $\alpha_{12} \leq
10^{-2}$. For a chameleon theory to satisfy this
bound we need the tests masses to have thin-shells. In
this scenario $d$ is small compared to the size of test-masses ($d <
R_{1}, R_{2}$) and so the previous formula for $\alpha_{12}$ does not
apply. When the mass of the chameleon inside the test masses, $m_{\phi}$, obeys $m_{\phi}d \gg 1$ (as is the case for
$\beta \geq 1$) we find that the $\phi$-force is $\alpha$ times the
strength of gravity, where $\alpha_{12}$ is\cite{doug}: $$ 5 \times
10^{-4}\left(\frac{M}{(0.1 \,
\mathrm{mm})^{-1}}\right)^{\frac{2(n+4)}{n+2}}\left(\frac{\sqrt{2}B\left(\frac{1}{2},\frac{1}{2}+\frac{1}{n}\right)}{\vert
n \vert d /\, 0.1 \, \mathrm{mm}}\right)^{\frac{2n}{n+2}}, $$ where
$B(p,q)$ is the beta function. We note that $\alpha$, as before, is
\emph{independent} of $\beta$. 

In this experiment a uniform $d_{\rm s} = 10\mu {\rm m}$ thick
BeCu membrane is placed between the test masses to shield
electromagnetic forces.  For $O(1)$ values of $\beta$ or $M \sim (0.1 \mathrm{mm})^{-1}$ this sheet does not have a
thin-shell and makes little difference to the analysis.  For slightly
larger values of $\beta$ however it will develop a thin-shell.  Taking the chameleon mass inside the sheet to be $m_{\rm s}$, the effect of this
membrane is to attenuate $\alpha_{12}$ by a factor of
$\exp(-m_{\rm s}d_{\rm s})$.  The larger $\beta$ becomes, the larger $m_{s}$ is and the less restrictive this bound becomes. 

The prospect that light scalar fields with couplings $\beta \gg 1$ could be allowed is
exciting and opens the door to many interesting and testable effects \cite{sc}. But to be taken seriously we must also consider bounds
coming from astrophysical constraints, such as the stability and
mass-radius relationship of white dwarfs and neutron stars as well as
bounds coming from BBN and the CMB anisotropies\cite{doug,chamcos}.  These bounds can be summarized as
requiring $\vert\beta \phi / M_{pl}\vert \leq 0.1$ over the whole
universe since the BBN epoch \cite{chamcos,doug}. In figure \ref{fig1} we show the bounds on chameleon models from the astrophysical and cosmological observations. 

In summary, 
chameleon fields strongly coupled to matter can be dark energy candidates and avoid at the same time stringent gravity experiments and cosmological bounds.
The reason is due to the  surprising result that the chameleon force
between two bodies with thin-shell is \emph{independent} of their
coupling to the field $\phi$, and that as a result the bounds on the
coupling, $\beta$, can be exponentially relaxed. 
When the chameleon plays the role of dark energy the strongest upper bounds on $\beta$ probably
come from particle colliders and $200 \mathrm{GeV} \leq
M_{pl}/\beta \leq 10^{15} \mathrm{GeV}$ is allowed for all $n$.
If $M_{pl}/\beta \sim 1 \mathrm{TeV}$ we might even hope to see
chameleon production at the LHC; 
Planned space-based tests such as
STEP, MICROSCOPE and SEE, promise improved precision
and, when $n > 0$ there is also still the possibility that WEP
violations in space can be stronger than the level already ruled out
by laboratory based experiments. 
\section*{Acknowledgments}
DFM and DJS acknowledge many discussions with  P. Brax, C. van de Bruck, A.C. Davis, J. Khoury and A. Weltman.
DFM and DJS are funded by Humboldt Foundation and STFC.


\section*{References}
\begin{thebibliography}{99}
%
\bibitem{c9}
  J.Khoury, A.Weltman,
  Phys.Rev.  D{\bf 69}, 044026 (2004);  
  Phys.Rev.Lett. {\bf 93}, 171104 (2004)
\bibitem{chamcos}
  P.Brax, C.van de Bruck, A.C.Davis, J.Khoury, A.Weltman,
  Phys.Rev. D{\bf 70}, 123518 (2004)
\bibitem{doug}
  D.F.Mota, D.J.Shaw,
  Phys.Rev.  D{\bf 75}, 063501 (2007);
  Phys.Rev.Lett.  {\bf 97}, 151102 (2006).
\bibitem{nelson}
  B.Feldman, A.E.Nelson,
  JHEP {\bf 0608}, 002 (2006)
\bibitem{sc}
  H.Gies, D.F.Mota, D.J.Shaw,
  Phys.Rev. D{\bf 77}, 025016 (2008); P.Brax, C.van de Bruck, A.C.Davis, D.F.Mota, D.Shaw,
  Phys.Rev.  D {\bf 76}, 124034 (2007);  P.Brax, C.van de Bruck, A.C.Davis, D.F.Mota,  D.J.Shaw,
  Phys.Rev.  D{\bf 76}, 085010 (2007); S.Das, N.Banerjee,
  arXiv:0803.3936 [gr-qc]; A.W.Brookfield et al.,
  \PRD {73}, 083515 (2006); A.E.Nelson, J.Walsh,
  arXiv:0802.0762 [hep-ph];
  P.Brax, C.van de Bruck, A.C.Davis,
  Phys.Rev.Lett. {\bf 99}, 121103 (2007);B.Li, J.D.Barrow, D.F.Mota,
  Phys. Rev.  D {\bf 76}, 104047 (2007);  D.A.Easson et al.,  JCAP {\bf 0802}, 010 (2008); S.Nojiri, S.D.Odintsov,
  Mod.\ Phys.\ Lett.\  A {\bf 19}, 1273 (2004); S.Capozziello, S.Tsujikawa,
  arXiv:0712.2268 [gr-qc]; D.F.Mota et al., Mon. Not. R. Astron. Soc. 382, 793-800 (2007),  arXiv:0708.0830 [astro-ph].; H.Wei, R.G.Cai,
  Phys.\ Rev.\  D {\bf 71}, 043504 (2005).
\end{thebibliography}
\end{document}